\documentclass[preprint,12pt]{aastex}
\usepackage{psfig}
\newcommand{\beq}{\begin{equation}}
\newcommand{\eeq}{\end{equation}}
\newcommand{\bea}{\begin{eqnarray}}
\newcommand{\eea}{\end{eqnarray}}
\newcommand{\bean}{\begin{eqnarray*}}
\newcommand{\eean}{\end{eqnarray*}}
\newcommand{\ba}{\begin{array}}
\newcommand{\ea}{\end{array}}
\newcommand{\bml}{\begin{mathletters}}
\newcommand{\eml}{\end{mathletters}}
\newcommand{\rem}[1]{{ }}

\bibpunct[,]{(}{)}{;}{a}{}{,}
\begin{document}
\title{Broadband modeling of GRB~021004}

\author{Jeremy S. Heyl\altaffilmark{1,2} \&
Rosalba Perna\altaffilmark{1,3}}

\altaffiltext{1}{Harvard-Smithsonian Center for Astrophysics, Cambridge, 
MA 02138; jheyl@cfa.harvard.edu, rperna@cfa.harvard.edu}
\altaffiltext{2}{Chandra Fellow} 
\altaffiltext{3}{Harvard Junior Fellow} 

\begin{abstract}
We present a broadband modeling of the afterglow of GRB 021004.  The
optical transient of this burst has been detected very early and
densely sampled in several bands. Its light curve shows significant
deviations from a simple power law.  We use the data from the X-ray to
the $I-$band gathered in the first month of observations, and examine
three models.  Two involve variations in the energy of the shock.  The
first (energy injection) allows only increases to the shock energy,
while the second (patchy shell) allows the energy to increase or
decrease.  In the final model (clumpy medium) the energy of the shock is
constant while the density varies.  While all three models reproduce
well the optical bands, the variable density model can best account
for the X-ray data, and the energy-injection model has the poorest
fit.  None of the models can account for the modest color variations observed 
during the first few days of the burst.
\end{abstract}

{

\section{Introduction}

The gamma-ray burst GRB 021004 was detected by HETE II at 12:06 UT on
the 4th of October 2002 \citep{GCN1565}. Observations
after about 9 minutes from the trigger revealed a fading optical
transient \citep{GCN1564}, which was densely sampled in several bands,
especially at early times. An X-ray counterpart has also been
reported by \citet{GCN1624}.

The afterglow of GRB 021004 has shown several unusual features.
Absorption lines from high velocity material have been detected in the
spectrum \citep{GCN1602,GCN1611,Moll02}, evidence of short time scale
variability has been presented \citep{Bers02b}, and the light curves
in various bands display significant deviations from a simple power
law decay, with several bumps and flattenings. A first, pronounced
bump around $10^3-10^4$ sec was modeled by \citet{Lazz02} with a
density contrast in the ambient medium density, while \citet{Koba02}
interpreted it as due to emission from the reverse
shock. \citet{Naka02} fitted the R band data and found that several
models provide good fits to the data, both with varying density (at
fixed energy) and with varying energy (at fixed density).  Determining
the appropriate physical model is crucial in order to understand what
the progenitors of GRBs are, and the large wealth of data available
for GRB 021004 in several bands allows detailed modeling.

In this {\em Letter}, we present a modeling of the light curve of GRB
021004 from the X-ray to the $I-$band.  We find that a model in
which the bumps are produced by density fluctuations (and not by
energy injection) best accounts for both the optical and the X-ray
data. We derive the density profile that best fits the data and
discuss its implications for GRB progenitor scenarios.

\section{Models}

To calculate the emission from the afterglow we have used the models
of \citet{Sari98} as extended by \citet{Naka02} for a varying density
or energy\footnote{Note that these models assume that all of the 
emission originates along the line of sight.  The emission actually 
comes from a surface of equal arrival times, so the actual lightcurve 
should be somewhat smoother than predicted by these models.}.
Although both the energy of the afterglow and the density
of the surrounding medium can vary, we will consider a GRB shock whose
energy changes while propagating through a wind with a constant
mass-loss rate or a constant-density ISM.  The calculations show that
the former model requires the energy of the shock to increase
monotonically so we will call it the ``the energy-injection model''.
The latter cases requires a shock whose energy varies up and down so
we will call it ``the patchy-shell model''.  The third model we
consider is a shock of constant energy traveling through medium with
varying density (the clumpy-medium model).

We refer the reader to \citet{Naka02} for the necessary equations, and we 
summarize their results here.
If $\nu_c < \nu_m$ (fast cooling), the flux density is proportional to
$\nu^{-1/2}$ between $\nu_c$ and $\nu_m$, to $\nu^{1/3}$ below $\nu_c$
and to $\nu^{-p/2}$ above $\nu_m$.  In both models for the assumed
values of the electron and magnetic field energy fractions of 0.1 and
0.01, fast cooling lasts for about the first thousand seconds after the
burst; therefore, only the data of \citet{GCN1564} lies in the
fast-cooling regime.  If $\nu_m < \nu_c$ (slow cooling), the flux
density goes as $\nu^{-(p-1)/2}$ between $\nu_m$ and $\nu_c$,
$\nu^{1/3}$ below $\nu_m$ and $\nu^{-p/2}$ above $\nu_c$.  $\nu_m$
passes through the optical and near infrared a few thousand seconds
after the burst.   On the other hand, changing the energy of the visible 
portion of the afterglow either through energy injection or a patchiness 
changes the flux density proportionally at all frequencies.

\section{Results}

We compare the three models with the broadband optical culled from the
GCN and the literature for the first day after the burst
\citep{GCN1564,GCN1566,GCN1573,GCN1591,GCN1594,GCN1606,GCN1615,Holl02,Pand02},
the X-ray data obtained by \citet{GCN1624} and reanalyzed ourselves,
and the later optical data from \citet{Bers02b}, \citet{Holl02} and
\citet{Pand02} .  We adjust either the energy injection or the density
of the surrounding medium to fit the $R$-band data.  The Chandra
observations of \citet{GCN1624} constrain the electron-injection
power-law exponent $p$ to be about $2.2$.  \citet{Math02} give a
redshift of 2.323 for the optical transient.  Assuming the
cosmographic parameters $\Omega_M=0.3, \Omega_\Lambda=0.7$ and
$H_0=70$~km~s$^{-1}$Mpc$^{-1}$ yields a luminosity distance of
$5.8\times 10^{28}$~cm to the gamma-ray burst.  This distance results
in an isotropic energy for the afterglow of about $9 \times
10^{52}$~ergs during the Chandra observation \cite{GCN1624}.  We
account for the cosmological time dilation and $k-$correction.

Because we varied either the energy of the forward shock or the
density of the surrounding medium to obtain a good fit to the
observations in the $R-$band and adjusted the extinction to fix the
average flux in the other bands, the key diagnostic between the two
models is that density variations only affect the flux between the
cooling frequency \citep{Lazz02,Naka02} and $\nu_m$ and are achromatic
in this regime.  Variations in the energy change the flux achromatically
throughout the near infrared, optical and X-ray regimes.
\begin{figure}
\plottwo{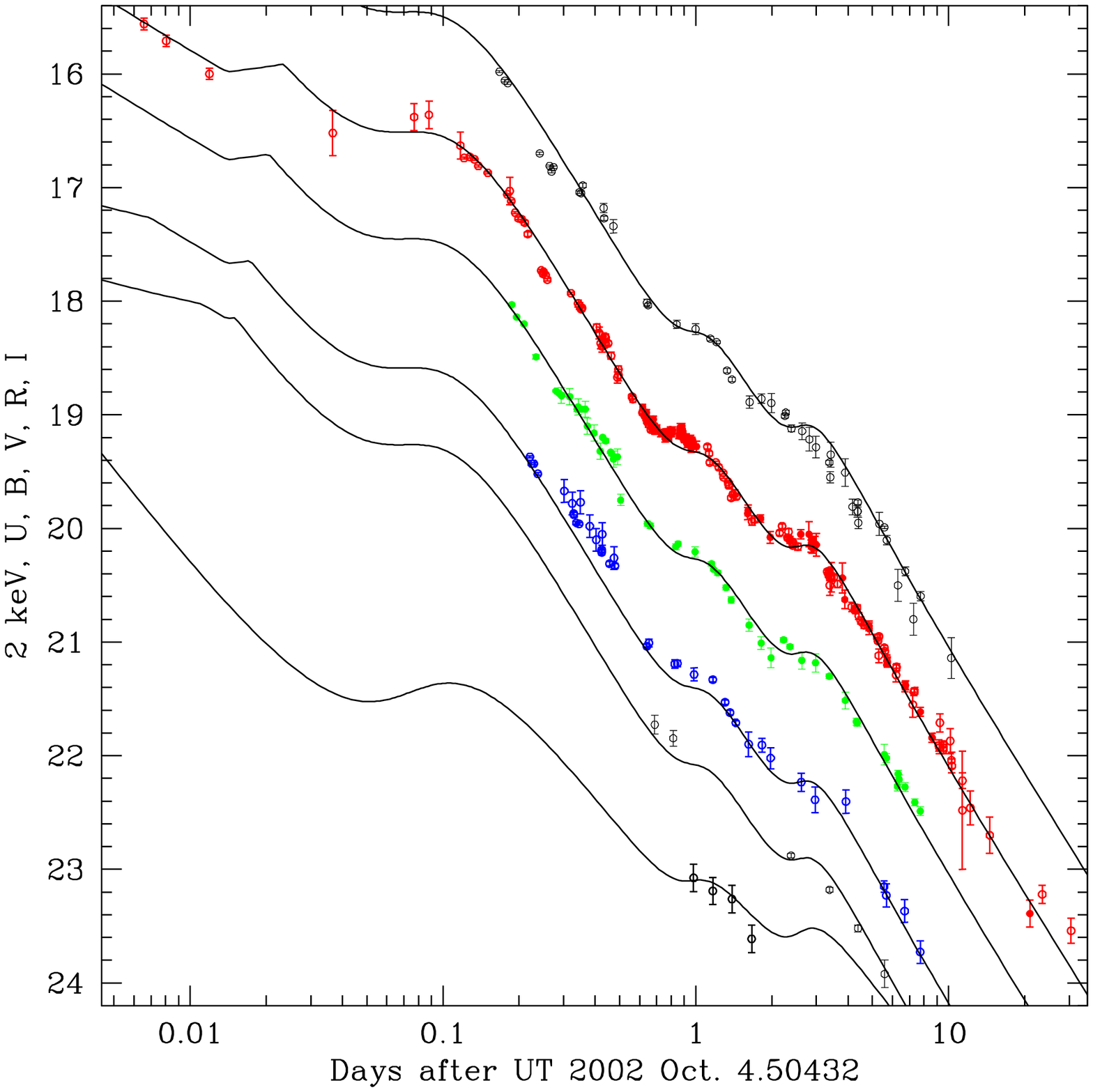}{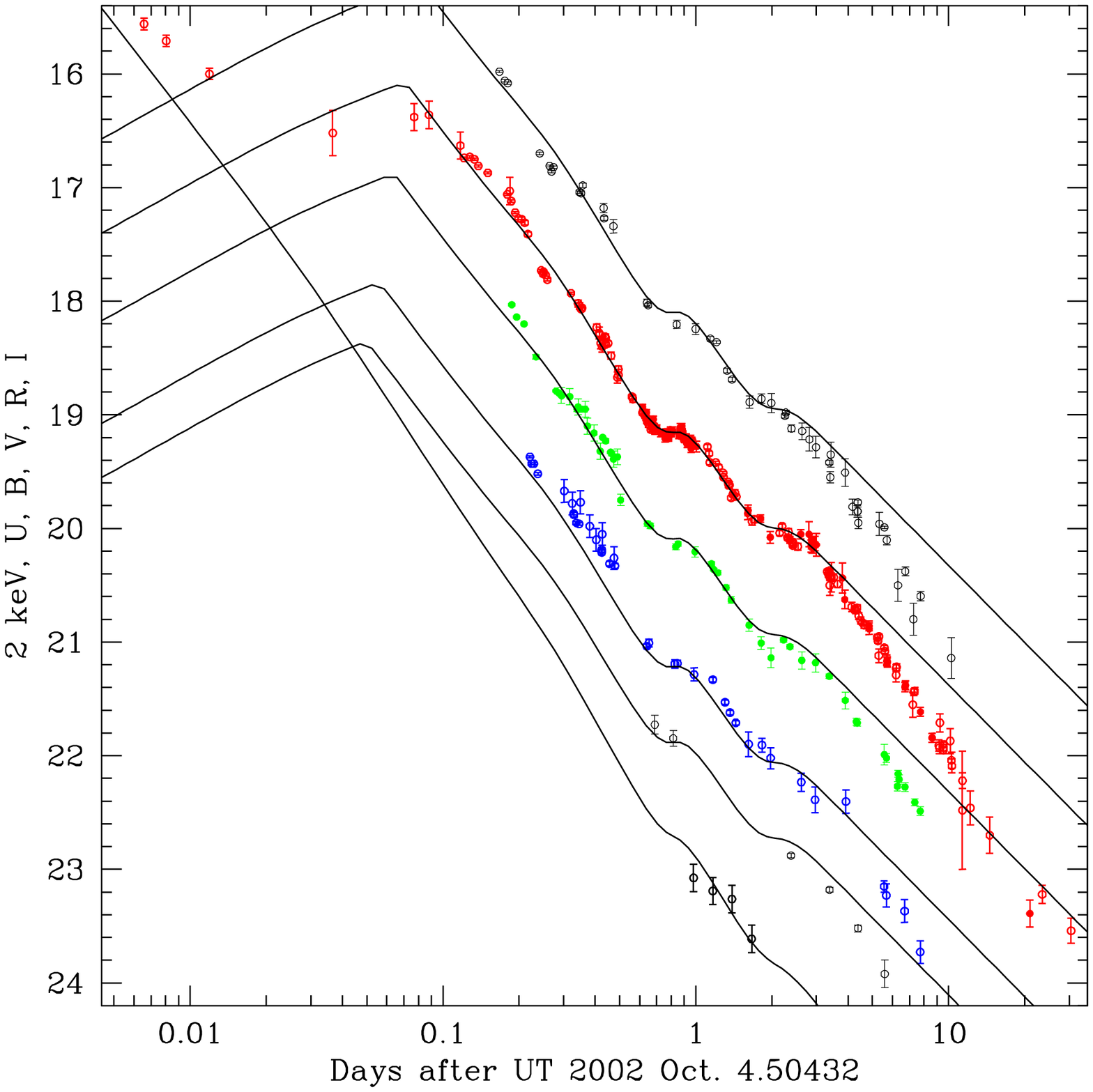}
\caption{The panels show the results of the two energy-variation models 
superimposed on the observations.   The left panel depicts 
the energy-injection model: the 
energy only increases as the shock propagates and the density drops 
as $r^{-2}$.   The right panel shows the patchy-shell model:  the density is constant 
and the energy varies up and down. From bottom to top, the curves trace the 
X-ray flux at 2~keV, $U+2$, $B+1$, $V+0.5$, $R$, and $I-0.5$.
The X-ray magnitude is defined using a zero point
800~Jy at 2~keV.}
\label{fig:broadband_energy}
\end{figure}
\begin{figure}
\plotone{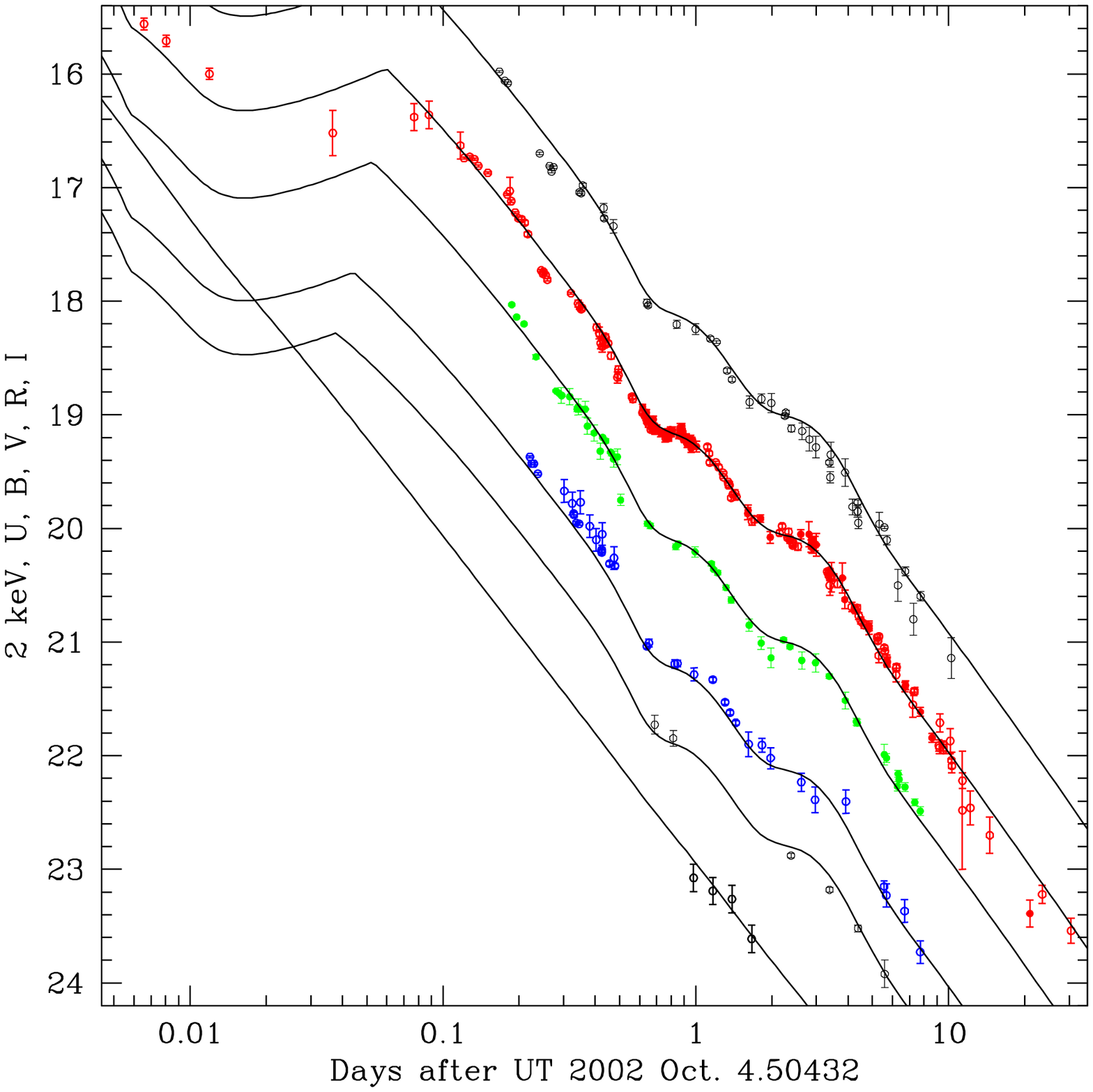}
\caption{The panel shows the results of the clumpy-medium model.  Here the
energy of the shock is constant but the density of the medium is allowed
to vary}
\label{fig:broadband_density}
\end{figure}

Fig.~\ref{fig:broadband_energy} compares the two energy-variation
models with the data from the first month after the burst.  For the
energy-injection model we have assumed a wind density profile that
corresponds to a mass-loss rate of $1.3 \times
10^{-6}$~M$_{\odot}$yr$^{-1}$, assuming a wind velocity of
1000~km~s$^{-1}$.  The local extinction is $A_{V'}=0.34$, where $V'$
is the local $V-$band.  The patchy-shell model has a constant density
medium with $n=0.1$~cm$^{-3}$ and a local extinction of $A_{V'}=0.33$.
Fig.~\ref{fig:broadband_density} depicts the clumpy-medium model.  In
this model, the isotropic energy of the ejecta is $8.9 \times
10^{52}$~ergs and the local extinction is $A_{V'}=0.33$.  This energy
agrees with the isotropic energy in gamma rays of the GRB itself
\citep{GCN1600}, $2.06 \times 10^{52}$~ergs if one assumes that
twenty-percent of the energy in the ejecta is converted to prompt
gamma rays \citep{Frai01}.  For all three models the Galactic value of
$A_V$ is taken to be 0.21 in the direction of the burst
\citep{Bers02b}.  We used the Milky-Way and LMC extinction models of
\citet{Pei92} to model the Galactic and host dust respectively.
For comparison \citet{Holl02} find $A_{V'}=0.26\pm0.04$ using Pei's SMC model.

All three models provide excellent fits to the optical and near
infrared data from one hour until about six days after the burst.  The
patchy-shell model fails to predict the flux during the first hour of
the burst and overpredicts the flux from the afterglow after six days.
The initial points may be explained by emission from a reverse shock
\citep[\protect{\em e.g.}][]{Pand02} or an increase in the density of
the medium near the progenitor.  \citet{Holl02} argue that the jet
break occurs at about six days, so because our models assume an
isotropic afterglow, the presence of a jet can account for this
discrepancy.  The energy-injection and clumpy-medium models can fit
the entire $R-$band light curve up to one month after the burst.

Fig.~\ref{fig:energy} depicts the required variation in the energy of
the visible portion of the fireball in the energy-variation models and
Fig.~\ref{fig:density} depicts the run of density in the clumpy-medium
model. In the clumpy-medium and patchy-shell models, the required
variations are modest so we are justified in neglecting the
contribution of reverse shocks to the emission \citep{Lazz02}.  The
energy-injection model requires the energy of the shock to increase by
a factor of forty, so the simple models of \citet{Naka02} may not be
applicable.  Note that, in the clumpy-medium model, the rise of the
density at small radii derives from fitting the early optical data
\citep{GCN1564} such a rise would not be required if that emission
were originated in the reverse shock as claimed by \citep{Pand02} or
if the early data were also in the slow cooling regime due to a
different choice of the parameters $\epsilon_e$ and $\epsilon_B$
(Lazzati priv. comm).  Although the transition from fast cooling to
slow cooling is dependent on $\epsilon_e$ and $\epsilon_B$, during
either regime changes in $\epsilon_e$ and $\epsilon_B$ are degenerate
with changes in the energy of the shock and the density of the medium.
\begin{figure}
\plotone{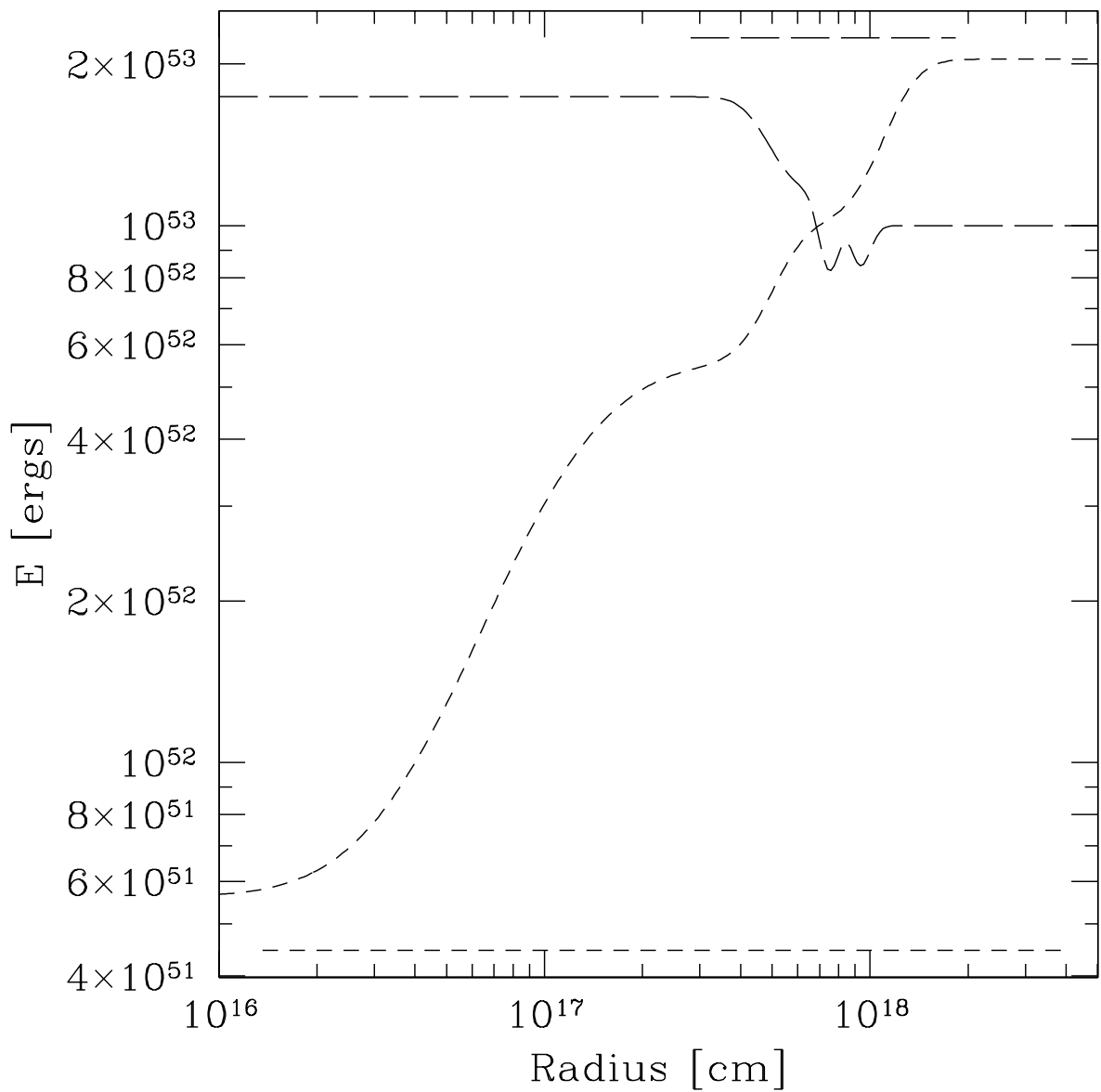}
\caption{The isotropic energy of the afterglow as a function
of radius for the energy-variation models.  The short-dashed curve follows
the energy-injection model, and the long-dashed curve follows the 
patchy-shell model.  The horizontal lines show the range of radii 
covered by the observations.}
\label{fig:energy}
\end{figure}

\begin{figure}
\plotone{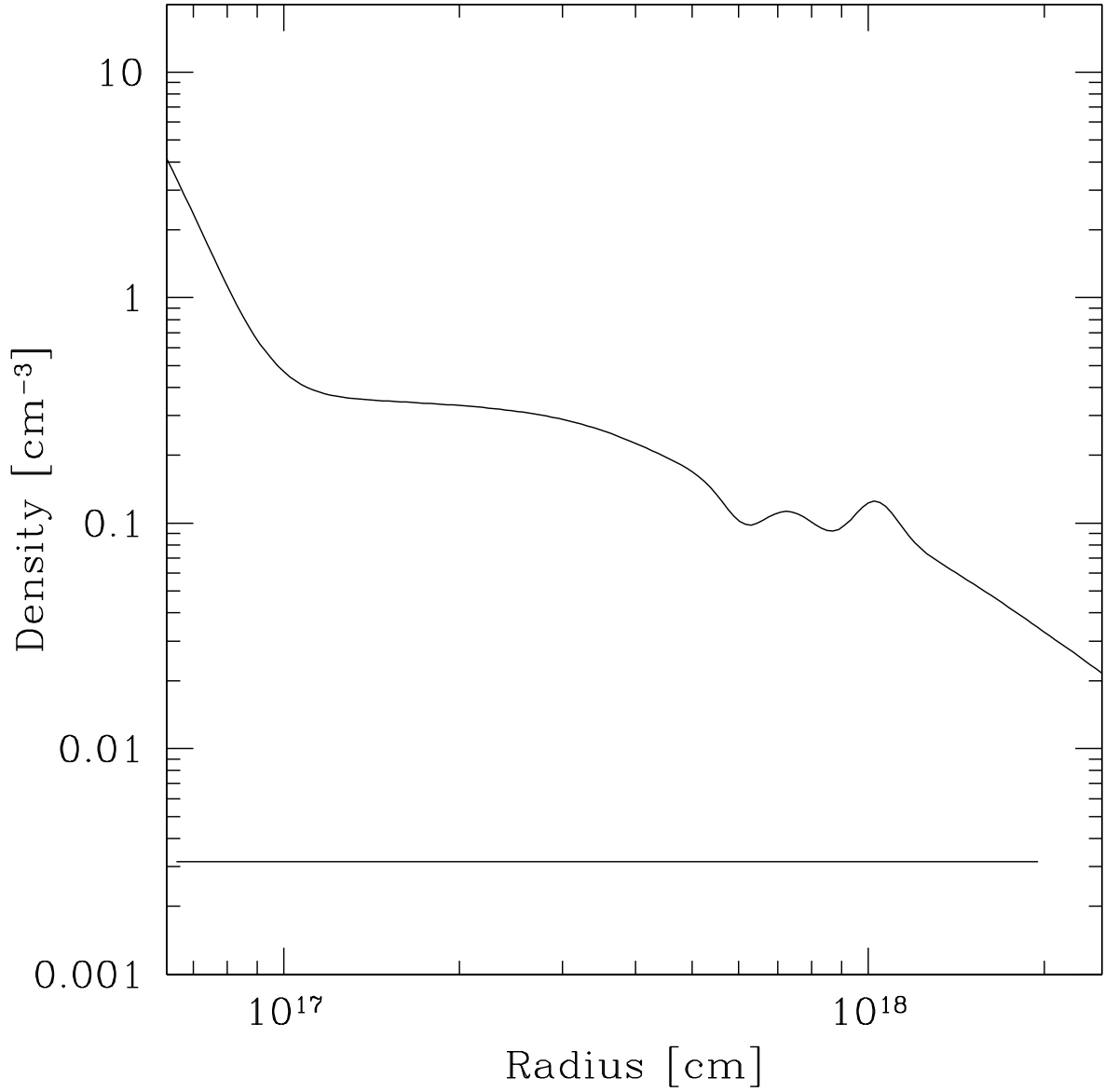}
\caption{The figure shows the density of the medium surrounding
the GRB progenitor as a function of radius.} 
\label{fig:density}
\end{figure}

\begin{figure}
\plotone{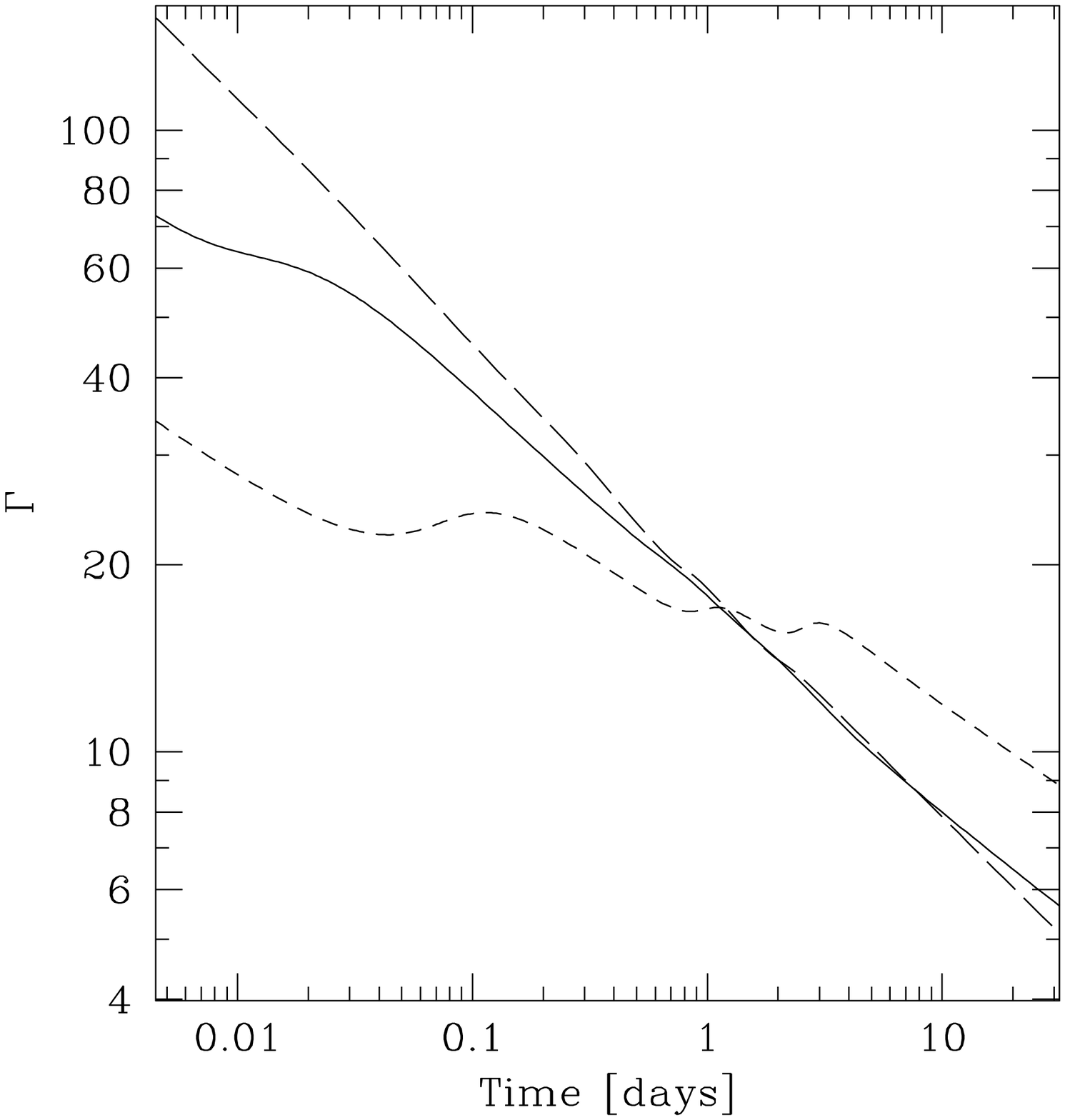}
\caption{The figure shows the value of the bulk Lorentz factor as a function
of the time after the burst. 
The solid curve traces the solution with clumpy-medium model,
the short-dashed lines follow the energy-injection result, and the long-dashed 
line gives the patchy-shell result.  The horizontal lines show the range of radii 
probed by the observations in each of the models.}
\label{fig:gamma}
\end{figure}

The largest difference between the models is at frequencies above
$\nu_c$.  The Chandra observations \citep{GCN1624} are contemporaneous
with optical and near infrared measurements \citep{Bers02b,Holl02,Pand02}.
The optical/NIR observations show significant variation from power-law
decay during this epoch.  The X-ray observations are consistent with
the power-law decay $t^{-1 \pm 0.2}$.  In the clumpy-medium model,
the X-ray flux follows a strict power-law decay during this epoch
because the cooling frequency lies below the X-rays.  On the other
hand, the energy-variation model results in complicated behavior in
the X-rays which is absent in the data.

We have reanalyzed the data obtained by \citet{GCN1624} to obtain
better constraints on the departures to the power-law evolution of the
flux in the X-ray band as they are crucial for discriminating among
the various models.
We looked at only the zeroth order image
which \citet{GCN1624} have argued is consistent with the HETG
spectrum.  We fit the ACIS spectrum for the entire observation with an
absorbed power-law energy distribution and obtained values consistent
with the results of \citet{GCN1624}.  We then compared the
distribution of photon arrival times for all those photons above the
median energy with those below the median energy.  Both these
distributions are consistent with the distribution of all of the
photon arrival times.  This indicates that there was little spectral
evolution during the observation, so we divided the observation into
four intervals each with the same number of counts and fit the
resulting spectrum with absorbed power-law energy distributions with
the power-law index and absorption column fixed, yielding the points
depicted in Figs.~\ref{fig:broadband_energy}
and~\ref{fig:broadband_density}.

To test our models robustly we used them to predict the distribution
of photon arrival times during the X-ray observations.  We compared
these distributions with the observed distribution using the
Kolmogorov-Smirnov statistic.  We found that the density-variation model
is preferred (K-S probability of 3\%) over the patchy-shell model
(K-S probability of 0.3\%).  The energy-injection model was highly disfavored
with a K-S probability of $2\times10^{-5}$.

If we examine in further detail, the color of the emission in the
optical and near infrared varies significantly from day to day
\citep{Bers02b,Math02}.  Because all of these bands lie
below the cooling frequency, both the density-variation and
energy-variation models cannot reproduce these color changes.  This is
apparent from Figs.~\ref{fig:broadband_energy}~and~\ref{fig:broadband_density}.

\section{Discussion}

We find that the broadband emission from GRB~021004 can be well
described by a relativistic fireball propagating into a clumpy
medium.  This model can account for the observed variability  in the
optical and near infrared while producing a power-law decay in the
X-rays.  Varying the energy in the observed portion of the fireball
results in a variation in the X-ray flux as well which was not seen. 

The patchy-shell model provides a poorer explanation.  In
this model, the mean energy of the observed portion of the fireball
can both increase or decrease \citep{Naka02} as it travels through a
constant density medium.  This model can account for the near infrared
through ultraviolet data, but it predicts a departure from power-law
decay in the X-rays.  It cannot account for the earliest data or for
the fluxes more than five days after the burst.  This time coincides
with the jet break found by \citet{Holl02}.  Because our models do not
account for jet emission, the jet provides a natural explanation for
the discrepancy.  In both the clumpy-medium and the patchy-shell
models $\Gamma \approx 10$ at this time.  Using the formula of
\citet{Frai01} gives a jet-opening angle of 0.09 consistent with the
value of $\Gamma$. This yields a total energy of the ejecta of
approximately $4 \times 10^{50}$~erg and $9\times 10^{49}$~erg in the
gamma-ray burst itself, somewhat less than the mean found by
\citet{Frai01}.

The least likely model is energy injection.  Because of the steep
decline in the second half of the first day, a wind density model is
required to fit the data in this scenario.  Without some significant
energy injection during this period, the model underpredicts the flux
later on.  Furthermore, the energy-injection model exhibits dramatic
evolution in the X-rays as well.  This was not seen.

We find that the afterglow of GRB~021004 was most likely produced in a
clumpy wind flowing into a low-density, typical ISM medium. The mean
density during most of the afterglow phase is about 0.1 cm$^{-3}$. If
this value is typical also at scales larger larger than those in which
the afterglow is produced, no variability in the absorption lines in
the spectra (associated with the host galaxy) should be expected
\citep{PL98}.  The typical density that we infer could be very well
typical of mergers of two compact objects as found in numerical
simulations \citep{PB02}, but in this case we do not expect a
particularly bumpy environment \citep[besides the small scale
fluctuations due to ISM turbulence, see][]{Wang00}, while this seems
to be the case for GRB~021004. Therefore, the association with the
collapse of a massive star is a more likely scenario.  A lumpy,
low-density background could suggest an explosion which occurred
inside the remnant of a previous supernova explosion, such as in the
Supranova model of \citep{VS98}.

Besides the magnitude of the density, the density profile itself in
the immediate environment of the burst can be be used to discern among
various progenitor models \citep[see \protect{\em e.g.}][]{Lazz01}. We
find that, if we fit the early optical data points within the same
forward shock model that is used to model the later
emission\footnote{This might not necessarily be the case according to
\citep{Pand02}.}, then the overall density profile resembles that of a
wind from a massive star, as one would expect in the collapsar
scenario \citet{MWH01}.  We have computed the mass loss rate that is
needed to produce the derived density assuming a wind velocity of 1000
km~s$^{-1}$.  Although the density that we find is somewhat lower than
that assumed by \citet{2001MNRAS.327..829R}, it is typical for
high-mass stars at subsolar metallicity \citep{2002ApJ...577..389K}.

GRB~021004 exhibited several color changes during its decay
\citep{Bers02b,Math02}.  None of the models that we have considered
can explain these changes.  Furthermore, it is impossible to reproduce
the color change within the broken-power-law spectral model of
\citet{Sari98} even by integrating over an heterogeneous fireball
\citep[\protect{\em e.g.}][]{Naka02}, unless the inhomogeneities are
sufficiently large to shift the values of $\nu_c$ or $\nu_m$ across
the bands of interest.  Because the latter possibility is unlikely, we
conclude that they are most likely either external to the fireball or
a result of the approximations used in current afterglow models.

\acknowledgements 

We would like to acknowledge useful discussions with David Bersier,
Kris Stanek, Matt Holman and Josh Winn, and we would like to thank
them for access to their data before release.  We also would like to
thank Davide Lazzati and the referee for useful comments.  JSH would
like to thank Bryan Gaensler and Antonella Fruscione for help with
analyzing the Chandra data.  He was supported by the Chandra
Postdoctoral Fellowship Award \#~PF0-10015 issued by the Chandra X-ray
Observatory Center, which is operated by the Smithsonian Astrophysical
Observatory for and on behalf of NASA under contract NAS8-39073.  We
used X-ray data obtained from the Chandra archive and reduced it using
CIAO and Sherpa.  We thank M. Sako and F. Harrison for requesting
director's discretionary time, Harvey Tananbaum for approving the
time, and the staff at the Chandra Science Center for skillfully
implementing the rapid TOO observation.

\bibliographystyle{apj}
\bibliography{mine,021004}

\begin{thebibliography}{31}
\expandafter\ifx\csname natexlab\endcsname\relax\def\natexlab#1{#1}\fi

\bibitem[{Barsukova {et~al.}(2002)Barsukova, Goranskij, Beskin,
  Plokhotnichenko, \& Pozanenko}]{GCN1606}
Barsukova, E.~A., Goranskij, V.~P., Beskin, G.~M., Plokhotnichenko, V.~L., \&
  Pozanenko, A.~S. 2002, GCN, 1606

\bibitem[{Bersier {et~al.}(2002)}]{Bers02b}
Bersier, D. {et~al.} 2002, \apjl, submitted (astro-ph/0211130)

\bibitem[{Fox(2002)}]{GCN1564}
Fox, D. 2002, GCN, 1564

\bibitem[{Frail {et~al.}(2001)}]{Frai01}
Frail, D.~A. {et~al.} 2001, \apjl, 562, L55

\bibitem[{Holland {et~al.}(2002)}]{Holl02}
Holland, S.~T. {et~al.} 2002, AJ, submitted (astro-ph/0211094)

\bibitem[{Klotz {et~al.}(2002)Klotz, Boer, \& Thuillot}]{GCN1615}
Klotz, A., Boer, M., \& Thuillot, W. 2002, GCN, 1615

\bibitem[{Kobayashi \& Zhang(2002)}]{Koba02}
Kobayashi, S. \& Zhang, B. 2002, \apjl, submitted (astro-ph/0210584)

\bibitem[{{Kudritzki}(2002)}]{2002ApJ...577..389K}
{Kudritzki}, R.~P. 2002, \apj, 577, 389

\bibitem[{Lamb {et~al.}(2002)}]{GCN1600}
Lamb, D. {et~al.} 2002, GCN, 1600

\bibitem[{Lazzati {et~al.}(2001)Lazzati, Perna, \& Ghisellini}]{Lazz01}
Lazzati, D., Perna, R., \& Ghisellini, G. 2001, \mnras, 325, L19

\bibitem[{Lazzati {et~al.}(2002)Lazzati, Rossi, Covino, Ghisellini, \&
  Malesani}]{Lazz02}
Lazzati, D., Rossi, E., Covino, S., Ghisellini, G., \& Malesani, D. 2002, A\&A,
  submitted (astro-ph/0210333)

\bibitem[{MacFadyen {et~al.}(2001)MacFadyen, Woosley, \& Heger}]{MWH01}
MacFadyen, A.~I., Woosley, S.~E., \& Heger, A. 2001, \apj, 550, 410

\bibitem[{Matheson {et~al.}(2002)}]{Math02}
Matheson, T. {et~al.} 2002, \apjl, submitted (astro-ph/0210403)

\bibitem[{Matsumoto {et~al.}(2002)Matsumoto, Kawabata, Ayani, Y.Urata, Yamaoka,
  \& Kawai}]{GCN1594}
Matsumoto, K., Kawabata, T., Ayani, K., Y.Urata, Yamaoka, H., \& Kawai, N.
  2002, GCN, 1594

\bibitem[{Mirabal {et~al.}(2002)Mirabal, Armstrong, Halpern, \& Kemp}]{GCN1602}
Mirabal, N., Armstrong, E.~K., Halpern, J.~P., \& Kemp, J. 2002, GCN, 1602

\bibitem[{Moller {et~al.}(2002)}]{Moll02}
Moller, P. {et~al.} 2002, A\&A, in press (astro-ph/0210654)

\bibitem[{Nakar {et~al.}(2002)Nakar, Piran, \& Granot}]{Naka02}
Nakar, E., Piran, T., \& Granot, J. 2002, New Astr., submitted
  (astro-ph/0210631)

\bibitem[{Oksanen {et~al.}(2002)Oksanen, Aho, Rivich, Rivich, West, \&
  Durig}]{GCN1591}
Oksanen, A., Aho, M., Rivich, K., Rivich, K., West, D., \& Durig, D. 2002, GCN,
  1591

\bibitem[{Pandey {et~al.}(2002)}]{Pand02}
Pandey, S.~B. {et~al.} 2002, Bull. Astr. Soc. India, submitted
  (astro-ph/0211108)

\bibitem[{Pei(1992)}]{Pei92}
Pei, Y.~C. 1992, \apj, 395, 130

\bibitem[{Perna \& Belczynski(2002)}]{PB02}
Perna, R. \& Belczynski, K. 2002, \apj, 570, 252

\bibitem[{Perna \& Loeb(1998)}]{PL98}
Perna, R. \& Loeb, A. 1998, \apj, 501, 467

\bibitem[{{Ramirez-Ruiz} {et~al.}(2001){Ramirez-Ruiz}, {Dray}, {Madau}, \&
  {Tout}}]{2001MNRAS.327..829R}
{Ramirez-Ruiz}, E., {Dray}, L.~M., {Madau}, P., \& {Tout}, C.~A. 2001, \mnras,
  327, 829

\bibitem[{Sako \& Harrison(2002)}]{GCN1624}
Sako, M. \& Harrison, F.~A. 2002, GCN, 1624

\bibitem[{Salamanca {et~al.}(2002)Salamanca, Rol, Wijers, Ellison, Kaper, \&
  Tanvir}]{GCN1611}
Salamanca, I., Rol, E., Wijers, R., Ellison, S., Kaper, L., \& Tanvir, N. 2002,
  GCN, 1611

\bibitem[{Sari {et~al.}(1998)Sari, Piran, \& Narayan}]{Sari98}
Sari, R., Piran, T., \& Narayan, R. 1998, \apjl, 497, L17

\bibitem[{Shirasaki {et~al.}(2002)}]{GCN1565}
Shirasaki, Y. {et~al.} 2002, GCN, 1565

\bibitem[{Uemura {et~al.}(2002)Uemura, Ishioka, Kato, \& Yamaoka}]{GCN1566}
Uemura, M., Ishioka, R., Kato, T., \& Yamaoka, H. 2002, GCN, 1566

\bibitem[{Vietri \& Stella(1998)}]{VS98}
Vietri, M. \& Stella, L. 1998, \apjl, 507, L45

\bibitem[{{Wang} \& {Loeb}(2000)}]{Wang00}
{Wang}, X. \& {Loeb}, A. 2000, \apj, 535, 788

\bibitem[{Weidinger {et~al.}(2002)}]{GCN1573}
Weidinger, M. {et~al.} 2002, GCN, 1573

\end{thebibliography}

\end{document}